\documentclass[jkps,reprint,twocolumn,showkeys]{revtex4}
\usepackage[pdftex]{graphicx}
\usepackage{amssymb}
\usepackage{amsmath}
\usepackage{bm}

\begin{document}
\title[]{Microscopic conductivity of passive films\\ on ferritic stainless steel for hydrogen fuel cells}

\author{Taemin \surname{Ahn}}
\author{Tae-Hwan \surname{Kim}}%
\email{taehwan@postech.ac.kr}
\affiliation{Department of Physics, Pohang University of Science and Technology (POSTECH), Pohang 37673, South Korea}


\begin{abstract}
Hydrogen fuel cells offer a clean and sustainable energy conversion solution. 
The bipolar separator plate, a critical component in fuel cells, plays a vital role in preventing reactant gas cross-contamination and facilitating efficient ion transport in a fuel cell. 
High chromium ferritic stainless steel with an artificially formed thin chromium oxide passive film has recently gained attention due to its superior electrical conductivity and corrosion resistance, making it a suitable material for separators.
In this study, we investigate the microscopic electrical conductivity of the intrinsic passive oxide film on such ferritic stainless steel. 
Through advanced surface characterization techniques such as current sensing atomic force microscopy and scanning tunneling microscopy/spectroscopy, we discover highly conductive regions within the film that vary depending on location. 
These findings provide valuable insights into the behavior of the passive oxide film in fuel cells. 
By understanding the microscopic electrical properties, we can enhance the design and performance of separator materials in hydrogen fuel cells. 
Ultimately, this research contributes to a broader understanding of separator materials and supports the wider application of hydrogen fuel cells.
\end{abstract}


\keywords{scanning tunneling microscopy, current sensing atomic force microscopy, hydrogen fuel cells, ferritic stainless steel, passive film, microscopic conductivity}

\pagenumbering{roman} 

\maketitle

\clearpage
\pagenumbering{arabic} 

\section{Introduction}
Hydrogen fuel cells are progressively becoming an appealing and sustainable technology, offering several benefits conducive to a transition towards a clean energy future~\cite{schafer2006FutureFuelCell, xu2020MassApplicationsReview}.
One key component of these cells is the bipolar separator plates.
These plates play an integral role in proton exchange membrane fuel cells, necessitating features like corrosion resistance, durability, and low contact resistance within the fuel cell stack~\cite{liu2022ReviewProtonExchange}. 
Although graphite and carbon composite materials are commonly used, stainless steel bipolar plates present a compact and potentially economical alternative, especially for portable and transport applications~\cite{yi2019LifetimePredictionModel, kraytsberg2007ReducedContactResistance}.

Stainless steel bipolar plates have advantageous properties such as low interfacial contact resistance, excellent corrosion resistance, high thermal conductivity, and low gas permeability~\cite{leng2020StainlessSteelBipolar, karimi2012ReviewMetallicBipolar}. 
Moreover, stainless steel offers desirable mechanical strength and formability, especially as a thin plate. 
However, these stainless steel plates pose a significant challenge due to the increased interfacial contact resistance between the stainless steel surface and the membrane electrode assembly layer~\cite{antunes2010CorrosionMetalBipolar, kraytsberg2007ReducedContactResistance, turan2011EffectManufacturingProcesses}.
This challenge is largely attributed to the semiconducting characteristics of the passive oxide film formed on the stainless steel surface under the fuel cell's typical operating conditions, usually a highly acidic environment~\cite{shaigan2021StandardizedTestingFrameworka, olsson2003PassiveFilmsStainless, decristofaro1997InfluenceTemperaturePassivation}.

To circumvent this challenge, a ferritic stainless steel has been developed featuring low interfacial contact resistance ($< 5$~m$\Omega \cdot$cm$^2$) and high corrosion resistance (corrosion current density $< 0.1$~$\mu$A$\cdot$cm$^{-2}$) through a pickling process followed by sophisticated passivation~\cite{kim2021posco}.
This stainless steel variant, which features a conducting Cr oxide passive film, shows high electrical conductivity, rendering it a promising material for bipolar plates~\cite{liu2022ReviewProtonExchange, leng2021ImprovementCorrosionResistance}.

However, oxide films on complex alloys such as stainless
steel are typically non-uniform.
This non-uniformity results in microscopic differences in electrical conductivity, which can hinder the overall performance. 
Hence, examining the microscopic conductivity traits becomes critically important from a practical application perspective. 
By understanding the relationship between the non-uniform conductivity and the other features of the oxide film, we could identify ways to enhance regions of the oxide film that demonstrate high electrical conductivity. 
In turn, this would enable us to achieve the higher electrical conductivity necessary for bipolar plates.

In this study, we aim to investigate the microscopic electrical conductivity of the chromium oxide passive film on the ferritic stainless steel bipolar plates.  
We have employed advanced microscopic characterization techniques such as current sensing atomic force microscopy (CSAFM)~\cite{afm_review, houze1996ImagingLocalElectrical, Hui2019} and scanning tunneling microscopy (STM)~\cite{stm_review, hansma1987ScanningTunnelingMicroscopy, feenstra1994ScanningTunnelingSpectroscopy} to unravel the underlying microscopic characteristics of the conducting passive film. 
Our findings have the potential to optimize the electrical performance of stainless steel bipolar plates, thereby facilitating their broader application in fuel cell technology.

\section{Experimental Details}
 Poss470FC, a commercially available variant of ferritic stainless steel, exhibits outstanding corrosion resistance and high electrical conductivity~\cite{kim2021posco}.
After a specialized chemical treatment, the resulting thin Cr oxide within a few nanometers from the surface significantly enhances Poss470FC's corrosion resistance and electrical conductivity, meeting the 2020 Department of Energy targets~\cite{doe2020}. 
To investigate the intrinsic properties of the passive film on the stainless steel, it is necessary to preserve the characteristic passive film under commercial conditions.
This work employed thin Poss470FC sheets cut into small samples, which were ultrasonically cleaned with ethanol to prevent surface contamination. 
This methodology guaranteed the relevance and validity of the findings regarding the inherent traits of the artificially formed Cr oxide on Poss470FC.
    
\begin{figure}[!th]
\centering
\includegraphics[width=8cm]{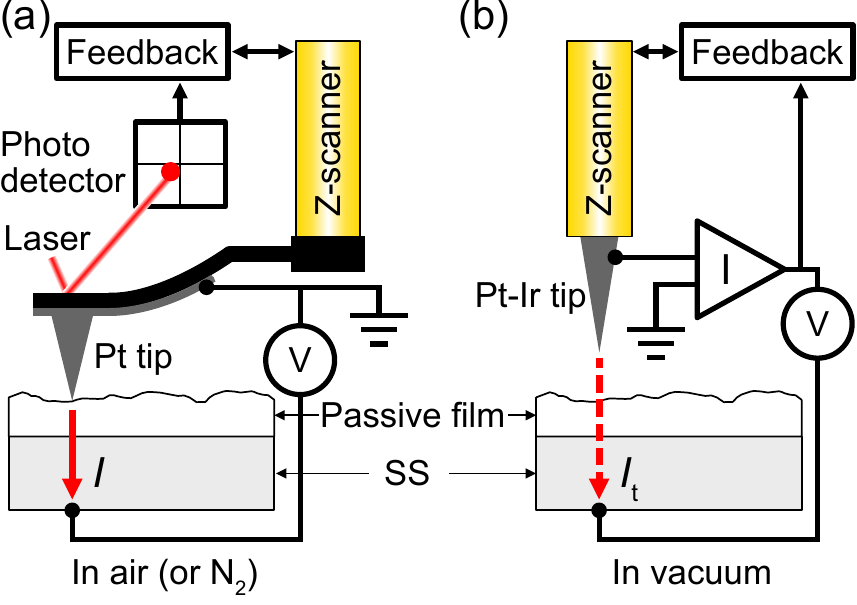}
\caption{\label{fig01}
(a) Schematic representation of current sensing atomic force microscopy (CSAFM) measurement of the passive film in air or a dry N$_2$ atmosphere.
(b) Schematic representation of scanning tunneling microscopy (STM) measurement in an ultrahigh vacuum (UHV) environment.
}
\end{figure}

In order to investigate the local conductivity of our samples, we employed a combination of CSAFM and STM (refer to the appendix for more details).
Incorporating current measurements with contact mode AFM imaging [Fig.~\ref{fig01}(a)], the CSAFM method, instrumental in exploring microscopic conductivity variations within resistive samples, was conducted using a commercial AFM (XE-100, Park Systems) equipped with solid Pt probe tips (25Pt300B, Rocky Mountain Nanotechnology). 
These tips were chosen for their ability to resist the degradation often associated with metal-coated probe tips~\cite{Jiang2019, weber2023SolidPlatinumNanoprobes, lantz1998CharacterizationTipsConducting}.
Despite its effectiveness, CSAFM comes with limitations like potential tip wear  and unavoidable disruptive influence on the sample surface~\cite{ji1998EffectsRelativeHumidity, stukalov2006RelativeHumidityControl, zitzler2002CapillaryForcesTapping, Ji2015, Jiang2019, Guo2006, jacobs2017MeasuringUnderstandingContact, sumaiya2020ImprovingReliabilityConductive}. 
To mitigate these limitations, we complemented our approach with STM measurements, which were performed under ultra-high vacuum (UHV) conditions ($P<1.0\times10^{-10}$~Torr) at room temperature using a home-built STM with an electrochemically etched Pt-Ir tip~\cite{cho2020SpecificStackingAngles}. 
STM, which involves positioning an atomically sharp metallic tip approximately 1~nm from the sample surface [Fig.~\ref{fig01}(b)], provides comprehensive atomic-scale topographic and electronic information without causing any damage to the samples.

\section{Results and Discussion}
\begin{figure}[th]
\centering
\includegraphics[width=8cm]{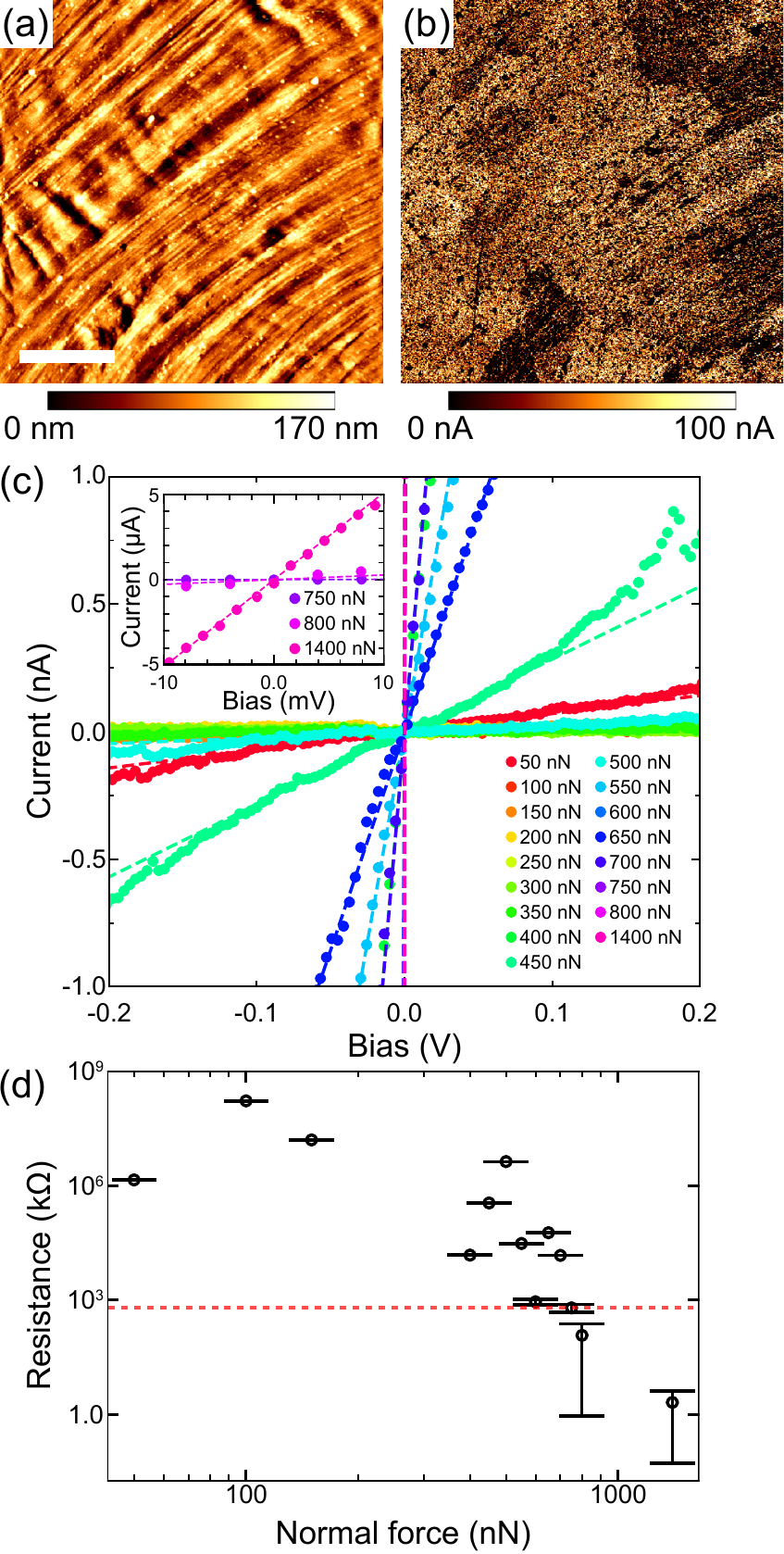}
\caption{\label{fig02}
(a) Contact mode CSAFM topographic image of the ferritic stainless steel and (b) its corresponding current map with a sample bias of $+0.5$~V and a tip normal force of 400~nN in air.
The scale bar is 10~$\mu$m.
(c) $I(V)$ curves obtained with varying tip normal force in the regions where higher current was obtained in the current map.
The inset better visualizes the 750, 800, 1400~nN data. 
(d) Estimated resistances derived from the linear regression of the reciprocal slopes of the measured $I(V)$ curves in the bias range of $\pm 0.1$~V.
For comparison, the red dashed line in the inset represents the resistance (650~k$\Omega$) obtained in a dry N$_2$ atmosphere with a normal force of 10~nN.
}
\end{figure}

To investigate the microscopic electrical conductivity of the passive film on ferritic stainless steel, CSAFM measurements were performed in ambient environments.
Figures~\ref{fig02}(a) and \ref{fig02}(b) depict the representative CSAFM topographic image and its corresponding current map of the ferritic stainless steel with a sample bias of $+0.5$~V, respectively.
The topographic image reveals typical features derived from the manufacturing process of the ferritic stainless steel.
Interestingly, the simultaneously obtained current map does not exhibit features correlated with topography, which suggests that the formation of the conductive passive film is not considerably affected by surface roughness.
Moreover, the current map uncovers the existence of two distinct regions.
These regions are differentiated by the measured current values, with areas of higher current predominantly surpassing those with lower current.
The prevalence of these higher current regions aligns well with the reported high electrical conductivity of the passive film on the ferritic stainless steel.

\begin{figure}[!ht]
\centering
\includegraphics[width=8cm]{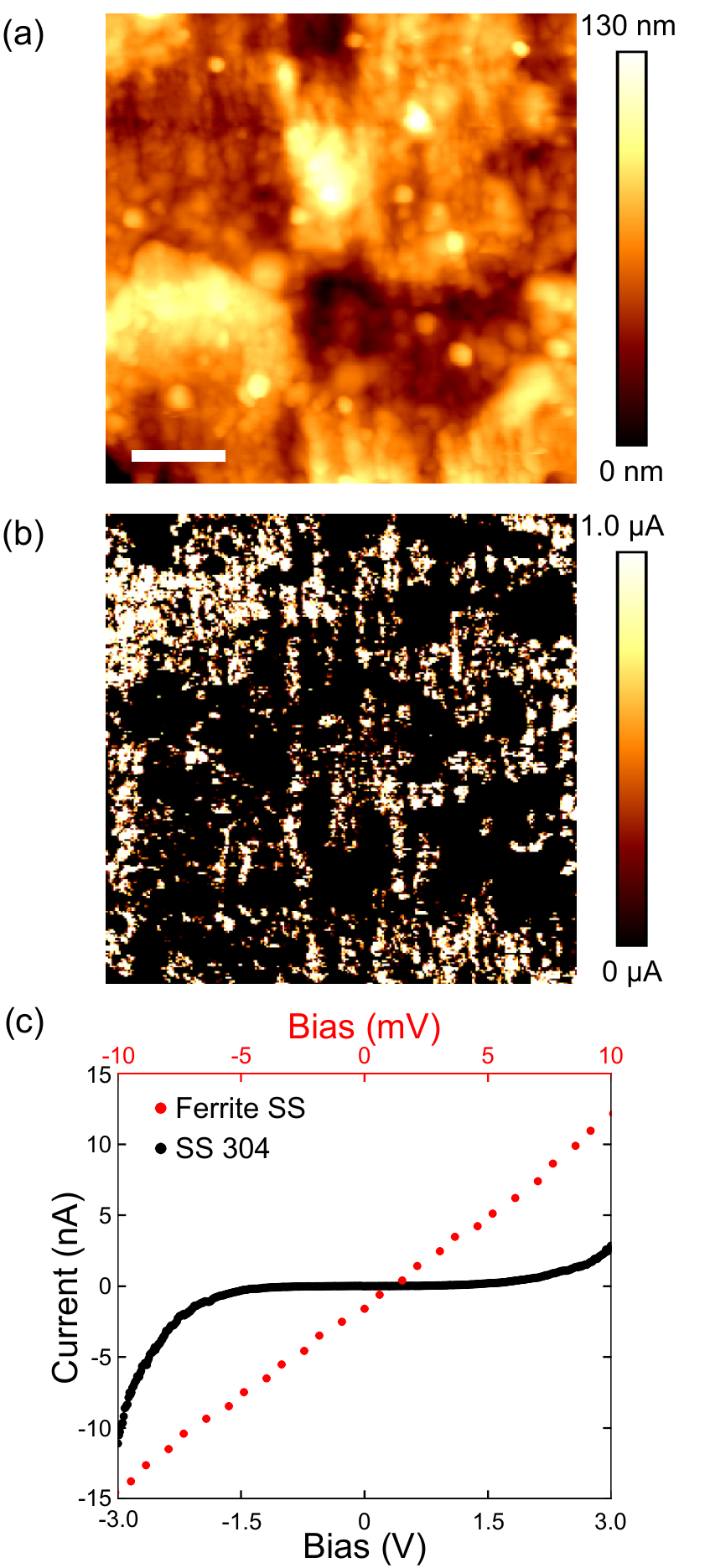}
\caption{\label{fig03}
(a) Contact mode CSAFM topographic image of the ferritic stainless steel and (b) its corresponding current map with a sample bias of $+0.1$~V and a normal force of 10~nN in a dry N$_2$ atmosphere.
The scale bar is 1~$\mu$m.
(c) Typical $I(V)$ curves obtained on the ferritic stainless steel (red dots) and stainless steel 304 (SS304, black dots) in the same dry N$_2$ atmosphere with a normal force of 10~nN.
}
\end{figure}

To definitively ascertain the ideal normal force for our ambient experiments, we systematically increased the normal force until we attained a stable $I(V)$ curve. 
We obtained local $I(V)$ curves from a region with elevated conductivity within the current map by adjusting the normal force from 50~nN to 1400~nN [Fig.~\ref{fig02}(c)].
The estimated resistance of the passive film, as a function of normal forces, was determined by calculating the reciprocal slope of the $I(V)$ curves within $\pm 0.1$~V, as shown in Fig.~\ref{fig02}(d).
In contrast to previous studies on passive films of other austenitic and ferritic phases~\cite{Souier2012, Lin2013, Guo2014}, our findings highlight the significantly enhanced conductivity (by at least a factor of 100) of the passive film on the ferritic stainless steel under comparable normal forces ($\sim 400$~nN).
This also lends credence to its superior electric conductivity compared to other stainless steel variants.
In general, the estimated resistance decreases with increasing normal force, which could be attributed to the inevitable water layer between the tip and sample under ambient conditions (refer to the appendix for more details)~\cite{Jiang2019, kremmer2003ModificationCharacterizationThin}.
However, it is important to acknowledge the potential risk of passive film fracture under excessive normal force,  as the estimated resistance was extraordinarily low above 800~nN.

\begin{figure}[!th]
\centering
\includegraphics[width=8cm]{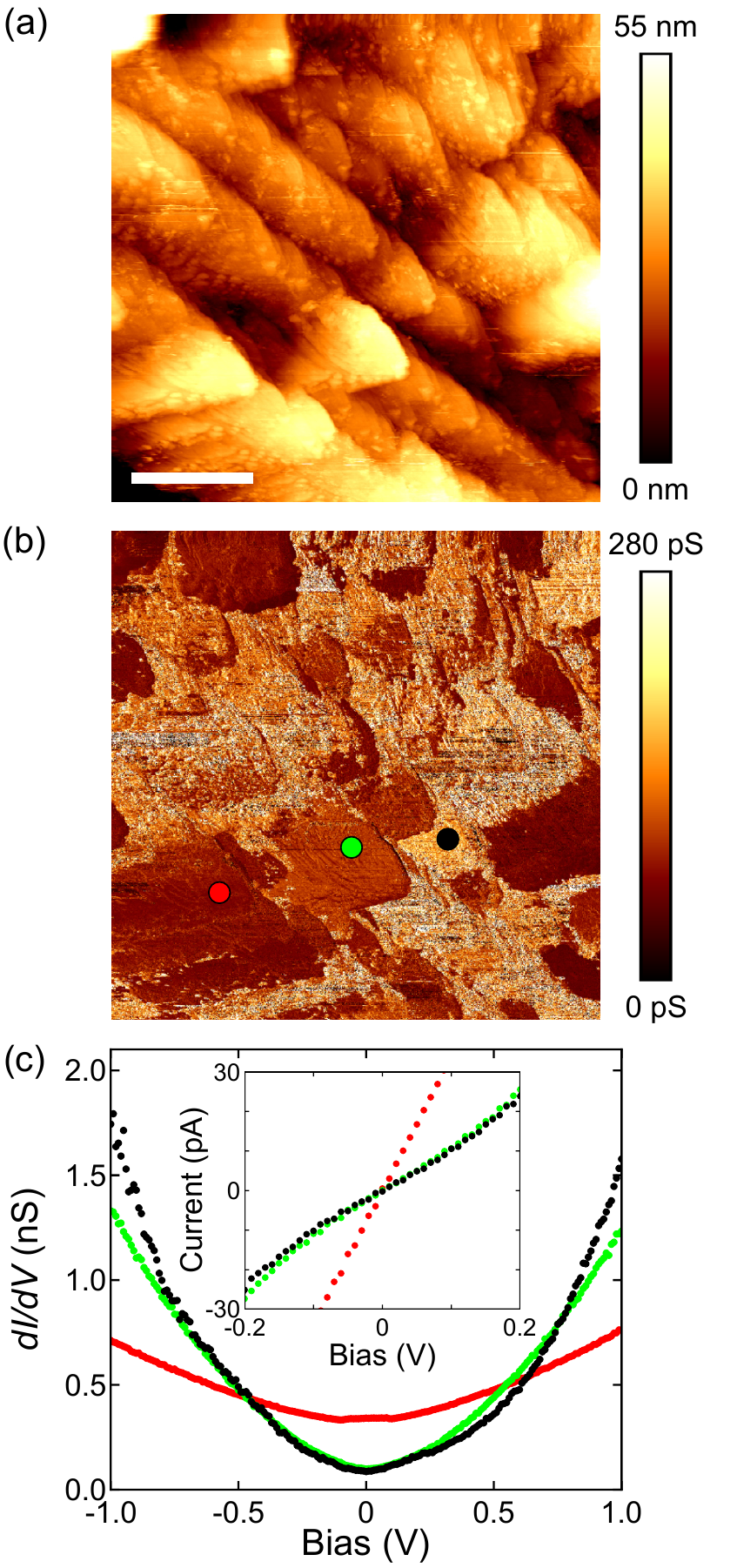}
\caption{\label{fig04}
(a) STM topographic image of the ferritic stainless steel and (b) its corresponding differential conductance ($dI/dV$) mapping image at a sample bias of $+1.0$~V. 
Imaging condition: $V_{\rm b}=+1.0$~V, $I_{\rm t}=50$~pA.
The scale bar is 400~nm.
(c) $dI/dV (V)$ and $I(V)$ curves (inset) obtained on the regions marked with different colored dots in (b).
}
\end{figure}

To avoid the potential risk of passive film fracture associated with excessive normal forces and current fluctuations induced by the presence of a water layer in ambient conditions, we carried out supplementary CSAFM measurements in a dry N$_2$ atmosphere~\cite{Jiang2019, Ji2015}. 
In contrast to the ambient conditions, we could successfully acquire reliable CSAFM current maps and $I(V)$ curves with the considerably reduced normal force in the dry atmosphere.
Figure~\ref{fig03} shows the CSAFM results of the ferritic stainless steel obtained with a normal force of only 10~nN (compared to 400--800~nN under the ambient conditions) and a sample bias of $+0.1$~V in the dry atmosphere.
Our findings were largely consistent with those observed under ambient conditions, except for the much lower normal force.
The current map also reveals the presence of two distinct conducting regions similar to those under the ambient conditions [Fig.~\ref{fig03}(b)].

For a direct comparison, we repeated the $I(V)$ measurement under the identical measurement conditions using a popular austenitic stainless steel variant (SS304).
Figure~\ref{fig03}(c) exhibits the ohmic behavior with low resistance of the passive film on the ferritic stainless steel, in contrast to the insulating characteristics of SS304.
To achieve a similar current range ($\pm 10$~nA), the $I(V)$ curve was captured within a sample bias range of merely $\pm 0.1$~V on the passive film of the ferritic stainless steel, whereas the curve on SS304 was obtained within $\pm 3$~V.
The estimated resistance was 650~k$\Omega$ on the passive film of the ferritic stainless steel, which is comparable to the resistance at a normal force of 750~nN in the ambient atmosphere [refer to the red dashed line in Fig.~\ref{fig02}(d)].

While we can prevent the passive film fracture in a dry atmosphere using a small normal force, we cannot overlook the mandatory contact resistance between the tip and sample.
This is due to the fact that the measured resistance invariably incorporates this contact resistance.
In response to this limitation of CSAFM, we further investigated the electronic properties of the passive film with STM under UHV conditions.
The quantum tunneling phenomenon between the tip and sample allows STM to provide local density of states (DOS) information and topography without mechanical contact, unlike CSAFM measurements [Fig.~\ref{fig01}(b)].

The ferritic stainless steel samples were transferred to our custom UHV STM chamber after being evacuated in the load-lock chamber overnight.
During this transition to the STM chamber, the sample was not subjected to any thermal treatment or annealing, ensuring to retain the intrinsic properties of the passive film.
Before the STM measurements, a metallic PtIr tip was routinely verified on an atomically clean Cu(100) surface.

Figures~\ref{fig04}(a) and (b) show the STM topography and its corresponding differential tunneling conductance ($dI/dV$) map, obtained at a sample bias of $+1.0$~V  and a tunneling current of 50~pA.
The STM image reveals more intricate but similar topographic features [Fig.~\ref{fig04}(a)], 
whereas the $dI/dV$ map shows three or more regions with different tunneling conductance intensities [Fig.~\ref{fig04}(b)].
The $dI/dV$ contrast represents local DOS differences of the oxide film surface obtained at 1.0~eV above the Fermi level.

Furthermore, we measured point scanning tunneling spectroscopy (STS) spectra on the regions showing distinct tunneling conductance [Fig.~\ref{fig04}(c)].
Three such regions are marked by red, green, and black dots in Fig.~\ref{fig04}(b).
Given that the local DOS is roughly proportional to the differential tunneling conductance in measured STS spectra~\cite{feenstra1994ScanningTunnelingSpectroscopy}, we inferred that the darker (brighter) region in the $dI/dV$ map [Fig.~\ref{fig04}(b)] presents higher (lower) local DOS at the Fermi energy ($V_{\rm b} = 0$~V) [Fig.~\ref{fig04}(c)]. 
These local DOS variations are likely due to subtle differences in the chemical composition of the oxide film.
Despite the spatially nonuniform tunneling conductance, we confirmed that the entire surface displays the metallic behavior without any electronic bandgap, thus reaffirming the superior electrical conductivity of the passive film. 

In sharp contrast to the CSAFM measurements, we observed a location dependence in the $dI/dV$ map and STS spectra in Fig.~\ref{fig04}.
This surprising discrepancy between the CSAFM and STM measurements could be attributed to the extreme sensitivity of STM measurements to the topmost surface, while CSAFM measurements encompass the entire thickness of the passive film.
Further investigation is required to gain a deeper understanding of this discrepancy between these complementary techniques.

\section{Summary}
In conclusion, our study offers a comprehensive microscopic investigation of the electrical conductivity of the chromium oxide passive film on the commercially available ferritic stainless steel, which has shown great promise for bipolar plate application in hydrogen fuel cells. 
Through the utilization of advanced surface characterization techniques such as CSAFM and STM/STS, we were able to distinctly identify both the remarkably high conductivity and the location-dependent conductance within the passive film.
These discoveries not only clarify the exceptional electrical conductivity of the material but also offer crucial insights that could be instrumental in enhancing the electrical performance of ferritic stainless steel bipolar plates. 
Although these findings mark significant progress, more research is needed to fully understand the location-dependent conductance and its implications for practical applications. 
Future work could explore this aspect further and investigate other potential materials for bipolar plate application. 
Ultimately, our research aids in the design and advancement of efficient separator materials for hydrogen fuel cells, thereby promoting the broader application of this sustainable energy technology in diverse fields.

\section{appendix}
\subsection{Current sensing atomic force microscopy (CSAFM)}
To explore the local conductivity of our ferritic stainless steel samples, we utilize CSAFM, which combines current measurements with contact mode AFM imaging.
In general, CSAFM operates under the standard AFM contact mode, using cantilevers coated with a conductive film~\cite{houze1996ImagingLocalElectrical, Hui2019}.
By integrating current measurements with contact mode AFM imaging, CSAFM serves as a powerful and effective method for investigating microscopic conductivity variations within resistive samples.
When a bias voltage is applied between the sample and the conducting cantilever, a current is induced [Fig.~1(a)], enabling us to obtain a spatially resolved conductivity image. 
CSAFM provides concurrent information on the spatial distribution of current and the sample topography with a constant cantilever load and bias voltage.
Furthermore, the measured current can be adjusted by varying the bias voltage and/or cantilever load.

For our specific experiments, we employ a commercial AFM (XE-100, Park Systems) with solid platinum (Pt) probe tips (25Pt300B, Rocky Mountain Nanotechnology), chosen for their ability to resist the degradation frequently encountered with metal-coated probe tips~\cite{Jiang2019, weber2023SolidPlatinumNanoprobes, lantz1998CharacterizationTipsConducting}. 
These probes, featuring a force constant of 18~N/m and a typical radius of less than 20~nm, allow for precise microscopic conductivity measurements.

We frequently encountered issues with unreliable current maps when using low normal forces.
This phenomenon can be attributed to the inevitable presence of a water layer between the metal probe tip and sample in ambient conditions.
The water layer results in the reduced electric field confinement near the tip-sample junction, causing an unstable current flow during the measurement process~\cite{Jiang2019, kremmer2003ModificationCharacterizationThin}.
To achieve stable current in CSAFM measurements under ambient conditions, it is imperative to apply a higher contact force to break the water layer~\cite{Jiang2019}.
In our study, we utilized a relatively high normal force of 400~nN in ambient conditions to guarantee highly reliable electrical contacts between the conducting tip and the sample during the process of current mapping~\cite{bietsch2000ElectricalTestingGold}.

Crucially, CSAFM allows for the simultaneous visualization of topography and current distribution, offering invaluable insights into the conductive characteristics of the passive film across different regions of the sample. 
However, as with any experimental method, CSAFM has limitations, such as the potential for tip wear and degradation, and possible disruptive influence on the sample surface~\cite{ji1998EffectsRelativeHumidity, stukalov2006RelativeHumidityControl, zitzler2002CapillaryForcesTapping, Ji2015, Jiang2019, Guo2006, jacobs2017MeasuringUnderstandingContact, sumaiya2020ImprovingReliabilityConductive}.
These factors can result in inconsistent or inaccurate measurements over time. 
To minimize these effects, we conducted our experiments under controlled humidity conditions, maintaining a relative humidity of less than 30\%. 
Additionally, we utilized N$_2$ gas flow to further decrease the humidity ($<9$\%)~\cite{Ji2015}.

For a more in-depth analysis, we measured representative current-voltage $I(V)$ curves in regions of the passive film exhibiting higher conductivity in the concurrently obtained current map and topographic image. 
Each point $I(V)$ curve was derived from an average of more than ten individual measurements taken at a fixed position. 
This approach ensured the reproducibility and consistency of the observed $I(V)$ characteristics, enhancing the reliability of our findings.

\subsection{Scanning tunneling microscopy (STM)}
To further enhance the analysis of our ferritic stainless steel samples, we expanded our methodology to include STM~\cite{hansma1987ScanningTunnelingMicroscopy}. 
This advanced technique provides atomic-level surface scrutiny,  mitigating the limitations associated with CSAFM. 
STM involves bringing an atomically sharp metallic tip into close proximity  with the sample surface (approximately 1~nm) [Fig.~1(b)].
When a bias voltage is applied between the tip and the sample, electrons tunnel through the vacuum barrier, generating a tunneling current that depends on the tip-sample distance, applied bias voltage, and the local DOS of the sample. 
We performed STM measurements in the constant current mode using an electrochemically etched Pt-Ir tip in a home-built STM under UHV conditions ($P<1.0\times10^{-10}$~Torr) at room temperature~\cite{cho2020SpecificStackingAngles}.

To refine our analysis, we STS, an advanced technique of STM~\cite{bert2015scanning, feenstra1994ScanningTunnelingSpectroscopy}. 
STS records the tunneling current response to varied bias voltages while maintaining a constant sample-tip distance. 
This non-invasive method enables us to obtain the local DOS of the sample, investigating its intrinsic electronic properties with atomic precision---an advantage over CSAFM. 
Combined with STM scanning mode, STS produces spatially resolved local DOS maps, providing detailed insights into the local electronic properties of the sample.
In this study, we obtained differential tunneling conductance ($dI/dV$) spectra using a lock-in amplifier that modulated the bias voltage by 30~mV at a frequency of 997~Hz.

While STM requires atomically clean and stable surfaces, strong vibration isolation, and high-performance electronics, it offers extraordinary spatial precision and unmatched atomic-level insights in terms of precision and detail. 
The combined use of STM and STS is instrumental in our investigation of the local conductivity of our ferritic stainless steel samples.

\begin{acknowledgments}
This work was supported by POSCO Steel \& Green Science (2019Y081).
\end{acknowledgments}


\end{document}